\documentclass{ws-procs9x6}
\usepackage{epsfig}

\def\trimmarks{}

\begin{document}

\title{UNINTEGRATED PARTON DENSITIES AND APPLICATIONS\footnote{Talk presented 
at the Ringberg Workshop, New Trends in HERA Physics, Schloss Ringberg, 
Germany, 28 Sept. -- 3 Oct. 2003.}}

\author{G\"OSTA GUSTAFSON}

\address{Dept. of Theor. Physics, Lund University,\\
S\"olvegatan 14A,
SE-22362 Lund, Sweden\\
E-mail: gosta@thep.lu.se}


\maketitle

\vspace*{-7cm}
\begin{flushright}
  LU-TP 03-55\\
  hep-ph/0312313\\
  November 2003
\end{flushright}
\vspace*{4.5cm}

\abstracts{
Different formalisms for unintegrated parton densities are discussed,
and some results and applications are presented.}
\vspace*{-1cm}

\section{Introduction}


Calculations based on \emph{collinear factorization} work very well in
many applications, including e.g.:

- DIS at large $Q^2$.

- High $p_\perp$ jets in $p\bar{p}$ collisions.

- Inclusive observables.

\noindent In these cases DGLAP evolution works, and $k_\perp$-ordered chains up to a 
hard subcollision or a highly virtual photon dominate the parton evolution.
The unintegrated parton density, $ \mathcal{F}(x,k_\perp^2)$, then satisfies 
the relations

\begin{equation}
F(x,Q^2) = \int^{Q^2} \frac{d k_\perp^2}{k_\perp^2} \mathcal{F}(x,k_\perp^2);
\,\,\,\,\,\,\,
\mathcal{F}(x,k_\perp^2) = \left. \frac{\partial F(x,Q^2)}{\partial 
\ln Q^2}\right|_{Q^2=k_\perp^2}.
\label{krullF}
\end{equation}

This formalism has, however, problems for observables which are sensitive to 
the $k_\perp$ of the ``active'' quark. Some examples are:

- Transverse momentum unbalance in 2-jet events.

- Heavy quark production.

- Forward jets at small $x_{Bj}$.

\noindent In many of these cases calculations,  which allow for 
one extra gluon, e.g. LO pQCD + parton showers or NLO DGLAP 
calculations, are able to give a good description of the data, but also these 
calculations have problems for observables which involve a large rapidity
separation. In the following I will discuss different formalisms for
$k_\perp$-factorization, non-$k_ \perp$-ordered evolution and some 
applications.

\section{Non-$k_ \perp$-ordered Evolution and $k_\perp$-Factorization}

At small $x$ and limited $Q^2$ non-$k_ \perp$-ordered chains give important
contributions. 
In a formalism based on $k_ \perp$-factorization the non-integrated pdfs 
$\mathcal{F}(x,k_\perp^2,Q^2)$ may depend on 2 scales, the $k_ \perp$ of
the active parton and the virtuality of the
photon or the hard scattering. The second scale is then related to a
limiting angle for the emissions, as discussed further below.
It is also important to remember that the gluon distribution is not an observable, 
and depends on the theoretical formalism.

\subsection{BFKL}

In the BFKL evolution\cite{BFKL}, accurate to leading $\log 1/x$, $\mathcal{F}(x,k_\perp^2)$ depends only on a single scale $k_ \perp^2$, and satisfies an integral
equation with a kernel $K(\mathbf{k}_\perp,\mathbf{k}_\perp')$:

\begin{equation}
\mathcal{F}(x,k_\perp^2) = \mathcal{F}_0 +\frac{3 \alpha_s}{\pi} 
\int \frac{d z}{z} \int d^2 k_\perp' K(\mathbf{k}_\perp,\mathbf{k}_\perp') 
\mathcal{F}(\frac{x}{z},k_\perp^{'2}).
\label{BFKL}
\end{equation}
We note that the dominant leading log behaviour originates from the $1/z$ pole 
in the splitting function. This leading order result has problems because
the NLO corrections are very large. There are e.g.
large effects from energy 
conservation, as demonstrated e.g. in MC studies by J. Andersen 
\emph{et al.}\cite{jeppe},
 which imply that analytic calculations often are unreliable.

The kernel $K$ in Eq.~(\ref{BFKL}) describes the emission of a quasi-real gluon
with transverse momentum $\mathbf{q}_\perp$ from a virtual link with 
momentum $\mathbf{k}_\perp'$, which after 
the emission gets momentum $\mathbf{k}_\perp = \mathbf{k}_\perp' - 
\mathbf{q}_\perp$. The kernel 
has the property that small values of $\mathbf{q}_\perp$ are suppressed.
Such emissions are compensated by virtual corrections, and in the BFKL
formalism this is taken into account by treating the links as Reggeized gluons.

We can compare this situation with $e^+e^-$-annihilation. Here the total cross 
section is determined by the lowest order diagram ($\alpha_s^0$). The
lowest order contribution to 3-jet events is $\mathcal{O}(\alpha_s)$, and to 
this order there are negative contributions to $\sigma_{2 \mathrm{jet}}$, such 
that $\sigma_{\mathrm{tot}}$ is approximately unchanged. In this case the 
gluon emissions can be treated with Sudakov form factors.

For a link in the BFKL chain the  $\mathcal{O}(\alpha_s)$ corrections
give a compensation for emissions for which $q_\perp \!< k_\perp$, 
$k_\perp'$. These soft emissions give no contribution to the inclusive 
cross section, i.e. to $F_2$. They must, however, be added for
exclusive final states, with appropriate Sudakov form factors. The net result 
of this is that downward steps in $k_\perp$ are suppressed by a factor
$k_\perp^2/k_\perp^{{'}2}$. In the relevant variable $\ln k_\perp^2$, this
corresponds to an exponential suppression allowing downward steps
within $\sim$ 1 unit in $\ln k_\perp^2$\cite{simple}.

\subsection{The CCFM model}

An evolution equation which interpolates between BFKL and DGLAP was 
formulated by Catani, Ciafaloni, Fiorani, and Marchesini\cite{CCFM}.
This formalism, the CCFM model, is based on a different 
separation between initial and final state radiation (denoted ISR and FSR 
respectively). Some soft emissions are 
included in the ISR, but this increase is compensated by 
``non-eikonal'' form factors. The ISR ladder satisfies the following 
constraint:

\vspace{3mm}
In the ISR emissions \emph{the colour order agrees with the order in energy} 
(or order in lightcone 
momentum $p_+ = p_0 + p_L$). (Thus all emissions which in colour order are 
followed by a more energetic one are treated as FSR.) With this constraint
colour coherence implies that \emph{this ordering also coincides with the 
ordering in angle, or rapidity}. 
\vspace{3mm}

This angular ordering implies that the 
emission angle for the last emission must be known, when a new step
is taken in the evolution. This  angle therefore 
appears as a limiting angle in the non-integrated distribution function used 
in the evolution. Thus the distribution $\mathcal{F}(x,k_\perp^2,\bar{q})$ 
depends on \emph{two} scales, $k_\perp^2$
and $\bar{q}$, where the limiting angle is specified by
$y_{\mathrm{limit}} = \ln(x M_p / \bar{q})$ (in the proton rest frame).

For exclusive final states, the final state radiation has to be added in 
appropriate kinematical regions. As the initial state radiation is ordered
in $p_+$ but not in $p_-$, also the final state radiation is unsymmetric
with respect to the initial proton and photon directions.

We can compare with the BFKL and DGLAP formalisms, where the parton 
distributions depend on a single scale. The initial state 
radiation is strongly ordered either in $q_+$ (for BFKL), or in $q_\perp$
(for DGLAP). In both cases this ordering
also implies a strong ordering in $y$. CCFM interpolates between the two 
regions at the cost of a more complicated formalism. 

That more emissions are included in the ISR implies that the average step is 
shorter. Therefore the angular ordering constraint becomes more
important and implies a strong dependence upon the scale $\bar{q}$.
This 
feature also implies that it has not been easy to implement the
CCFM model in an event generator, but such a program, \textsc{Cascade} by 
Hannes Jung\cite{CASCADE}, is now available. The CCFM model does not include
quark links in the parton chains. The original model, and also the first 
version
of the \textsc{Cascade} program, referred to as JS, also included only the
singular term $\propto 1/z$ in the splitting function. To include the
non-singular terms is not straight forward, but one possible solution is
implemented in a new fit, called J2003\cite{j2003}. In this fit set 1 
includes only the singular 
terms, while in set 2 also the non-singular terms in the splitting function 
are included.

\subsection{The Linked Dipole Chain Model}

The Linked Dipole Chain Model, LDC\cite{LDC}, is a reformulation and generalisation
of the CCFM model. It is based on a different separation between initial and 
final state radiation. The ISR is ordered in both $q_+$ and $q_-$, and
satisfies the constraint 
$q_{\perp i} > \min(k_{\perp i},k_{\perp i-1})$.
Softer emissions are treated as final state emissions. In this respect the 
LDC formalism is more similar to BFKL. We note that the ordering in
$q_+$ and $q_-$ also implies an ordering in angle. An important property
is also that the parton chain is fully symmetric with respect to the two ends 
of the chain.

The fact that fewer emissions are treated as ISR implies that a single
chain in LDC corresponds to the collective contributions from 
several possible chains in CCFM. Now it
turns out that summing over all possible emissions in CCFM, the non-eikonal
form factors exactly cancel. The result is a simple evolution equation
in terms of a \emph{single scale} unintegrated density function
$\mathcal{F}(x, k_\perp^2)$. In the MC implementation it is, however, 
also possible to 
add an angular cut and thus obtain results for a two-scale distribution.

We note that to leading order the LDC and CCFM
formalisms give the same result for the integrated structure functions.
The parton chains and the unintegrated distributions differ, however, and only
after addition of final state emissions in the different 
relevant kinematic regions do
the two formalisms also give the same result for exclusive final parton states
(to leading order).

Thus the LDC formalism results in a much simplified evolution equation. 
Other merits of the formalism include:

- It contains the same chains as in DGLAP for $Q^2$ large, which makes it 
easier to interprete the differences between large and small $Q^2$.

- There is a natural generalization to include subleading terms, 
e.g. quark links, non-singular terms in the splitting functions, and
a running $\alpha_s$.

- It is suitable for implementation in a MC, and the event generator LDCMC
is produced by L\"onnblad and Kharraziha\cite{LDCMC}.

- The MC gives very good fits to experimental $F_2$ data,
and it also agrees
well with MRST and CTEQ results for the \emph{integrated} gluon 
distribution\cite{gluon}.

\subsection{Other formalisms for unintegrated parton densities}

A different formalism, which also interpolates smoothly between BFKL and DGLAP,
was formulated by Kwieci\'nski, Martin, and Sta\'sto (KMS)\cite{KMS}. This
formulation is based on a single scale evolution equation. The contributions 
from $k_\perp^2 > Q^2$ are neglected, and thus the 
gluon distribution satisfies Eq.~(\ref{krullF}).
This formalism has been  further developed
by Kimber, Martin and Ryskin (KMR)\cite{KMR}. In their approach 
the single scale KMS evolution is used, but
an angular constraint is applied for the last step. The result is therefore a 
density distribution, which depends on two scales. Since the underlying
KMS evolution does not include the soft initial state radiation
in the CCFM model, it also takes fewer and larger steps, which implies that
the dependence on the
limiting angle is significantly smaller than for the CCFM formalism.

\section{Comparison between results from different formalisms}

\begin{figure}
\begin{center}
\epsfig{figure=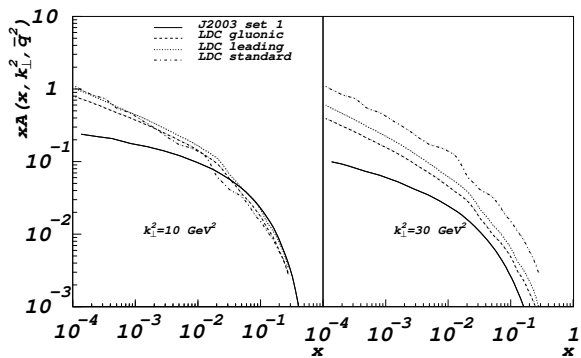,width=8.4cm}
\end{center}
\caption{Comparison of different sets of unintegrated gluon densities at
scale $\bar{q}=10$ GeV. For the notations, see the main text.}
\label{unintegratedpdf}
\end{figure}

\begin{figure}
\begin{minipage}{13cm}
\epsfig{figure=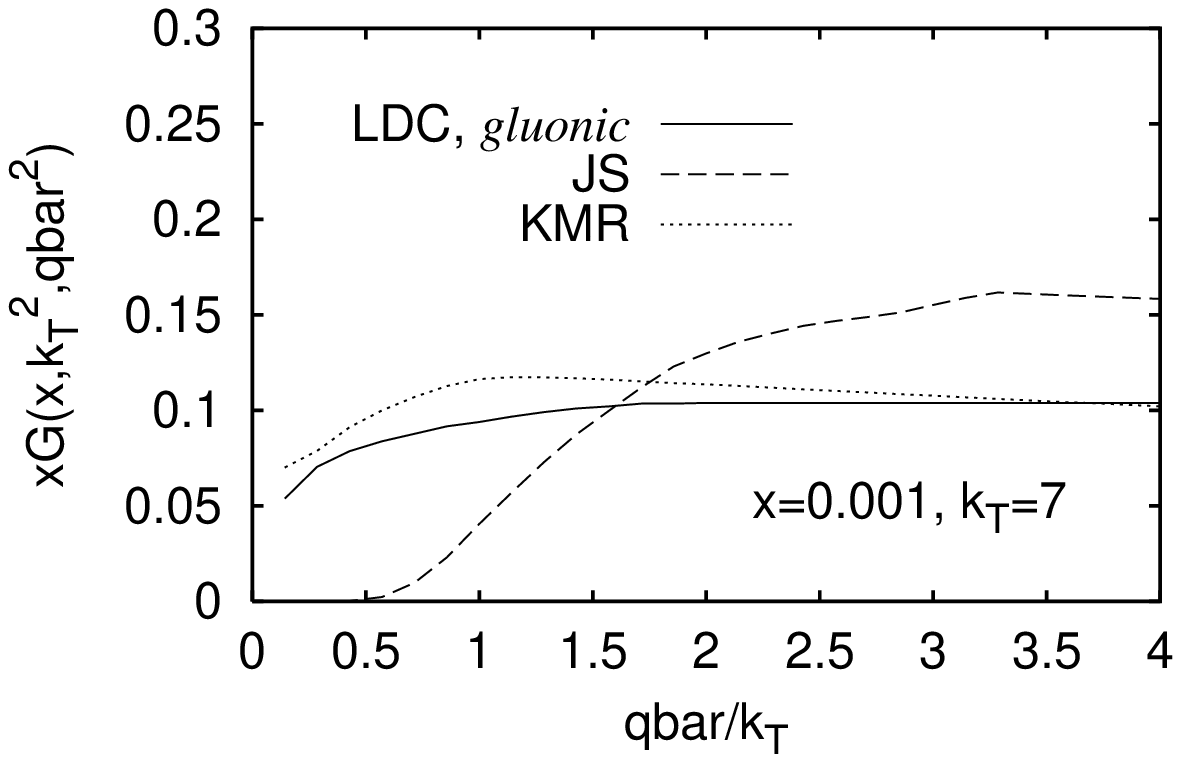,width=6cm,height=4.5cm}\hspace{-4mm}
\epsfig{figure=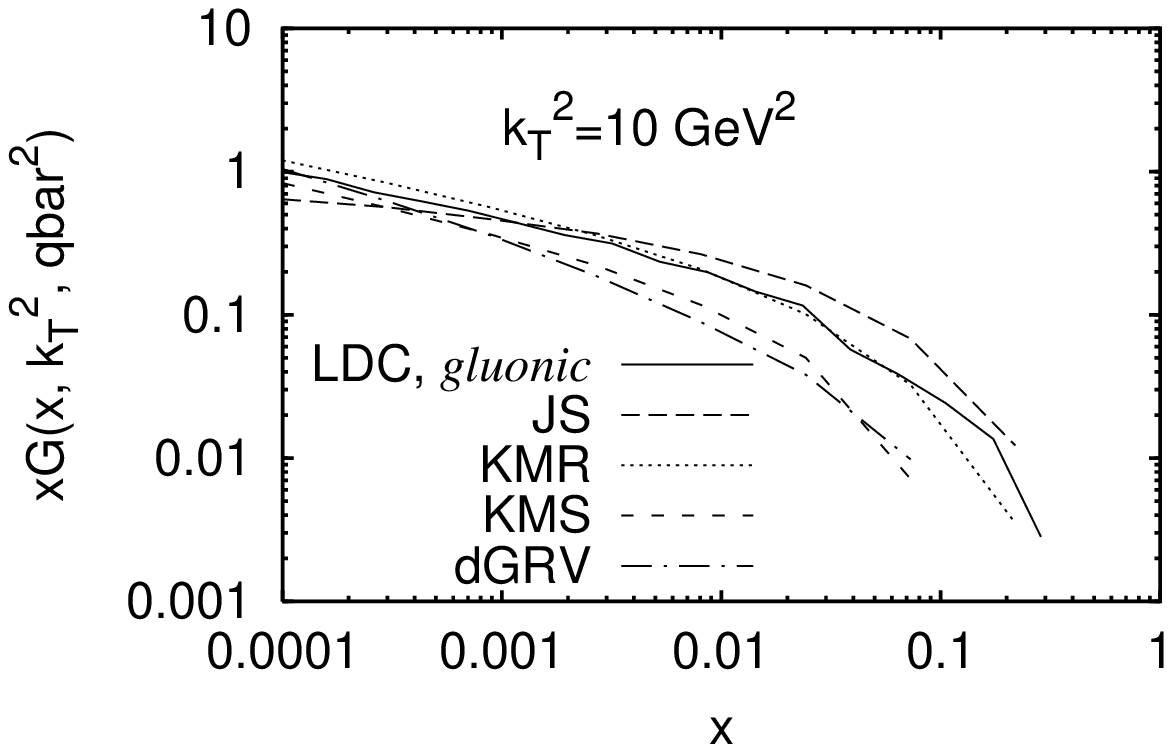,width=6cm,height=4.5cm}
\end{minipage}
\caption{The LDC \emph{gluonic} unintegrated gluon density
compared to the results of JS  and KMR. \emph{Left}: As function of 
$\bar{q}/k_\perp$ for fixed $x$ and $k_\perp$. 
\emph{Right}: As function of $x$ for $k_\perp^2 = 10\,\, \mathrm{GeV}^2$ and 
$\bar{q}= 2 k_\perp $. Also shown are here the single scale results from KMS 
and the derivative of the GRV fit.}
\label{unint}
\end{figure}

We here want to compare the parton densities obtained in different formalisms,
and also study the effects of the non-singular terms in the splitting 
functions and of quark links in the evolution chains.
Some results are shown in Fig.~\ref{unintegratedpdf}\cite{gluon} 
\cite{Hannesprivat}. Among the LDC
fits the result denoted 
\emph{standard} contains both quark links and non-singular splitting terms, 
\emph{gluonic} 
contains only gluon links, and \emph{leading} only singular terms. 
(In the fit \emph{gluonic-2}, refered to in the next section, the power of
$(1-x)$ in the input gluon density is changed to 7 instead of its value 4
in \emph{gluonic}.) In all cases
the input distributions for $k_\perp^2 = Q_0^2$ are adjusted to give good fits 
to experimental $F_2$ data. The CCFM result J2003 
set 1 contains only gluons and only the $1/z$ pole in the
splitting function. We see 
that the effect of quarks and non-singular terms become larger for larger 
$k_\perp$.
The sensitivity to the scale $\bar{q}$ 
is illustrated in Fig.~\ref{unint}a, which
shows the gluon
density $\mathcal{F}(x,k_\perp^2,\bar{q})$ as function of $\bar{q}/k_\perp$ 
for fixed $k_\perp$. As discussed above, in the CCFM 
formalism the density is very sensitive to the $\bar{q}$ scale, and 
varies strongly for $\bar{q}/k_\perp$ between 1 and 2.

It is, however, interesting to note that observable cross sections
differ much less than the parton densities, as seen in the next section. 
The reason is that the distributions differ
in particular for $\bar{q}<k_\perp$, while in a hard subcollision the
dominant contributions are obtained for 
$k_\perp^2 \approx \bar{q}^2/4$ \cite{gluon}. In Fig.~\ref{unint}b
we see that the results are indeed not so different for 
$\bar{q}= 2 k_\perp$. In this figure also the single scale result from KMS and 
the derivative of the GRV result (denoted dGRV) are included.

\section{Applications}

\subsection{Heavy quark production}

\begin{figure}
\begin{minipage}{13cm}
\epsfig{figure=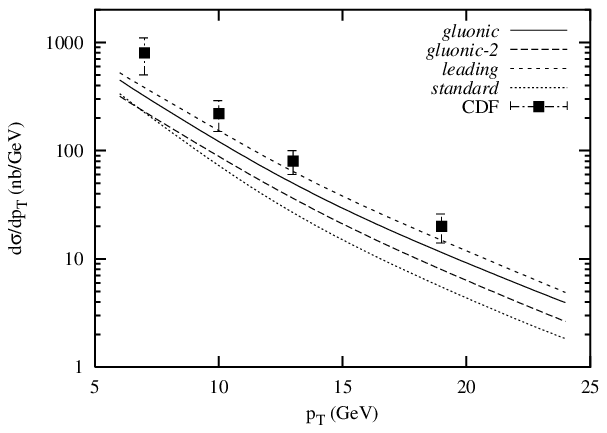,width=6.3cm}
\epsfig{figure=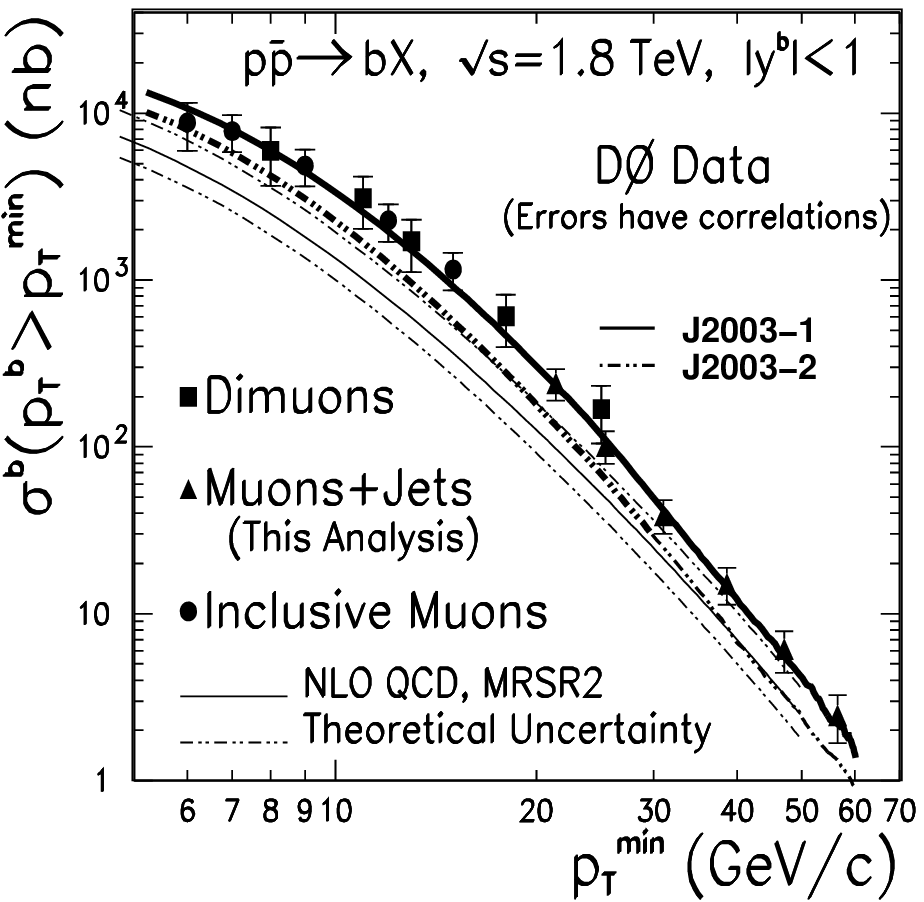,width=5.0cm}
\end{minipage}
\caption[bla]{\emph{Left}: Production of $b$-quarks at CDF\cite{CDFb} 
compared to 
results from the LDC model\cite{LDCb}. \emph{Right}: D0 data\cite{D0b} 
compared with CCFM results\cite{Hannesb}. Here also a NLO QCD result is 
included. For notations, see the main text.}
\label{bprod}
\end{figure}

Applications of the $k_\perp$-factorization formalism to $b$ quark production
at the Tevatron are shown in Fig.~\ref{bprod}. We see that for both the LDC 
and the CCFM models the fits with only the leading term reproduce
the data best, while the other fits, which should be expected to be
more accurate, although being significantly better than the NLO QCD
result, do not give equally good fits. We should note, however, that $b$ 
production is also well reproduced by collinear factorization plus
parton showers, as implemented in the \textsc{Pythia} event 
generator\cite{PYTHIAb} \cite{Fieldb}.

\subsection{Forward jets}

\begin{figure}
\begin{minipage}{13cm}
\epsfig{figure=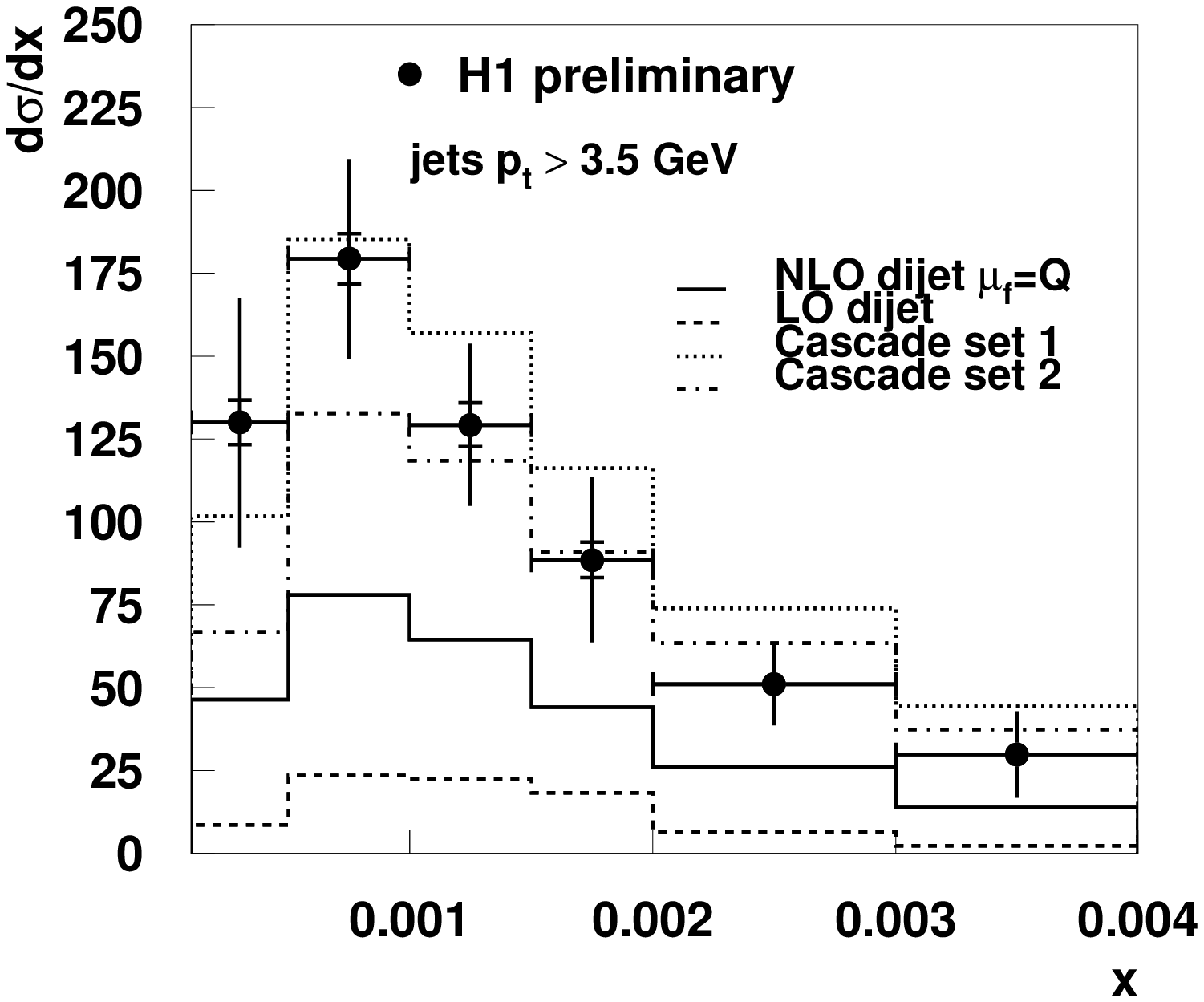,width=6cm}\hspace{-4mm}
\epsfig{figure=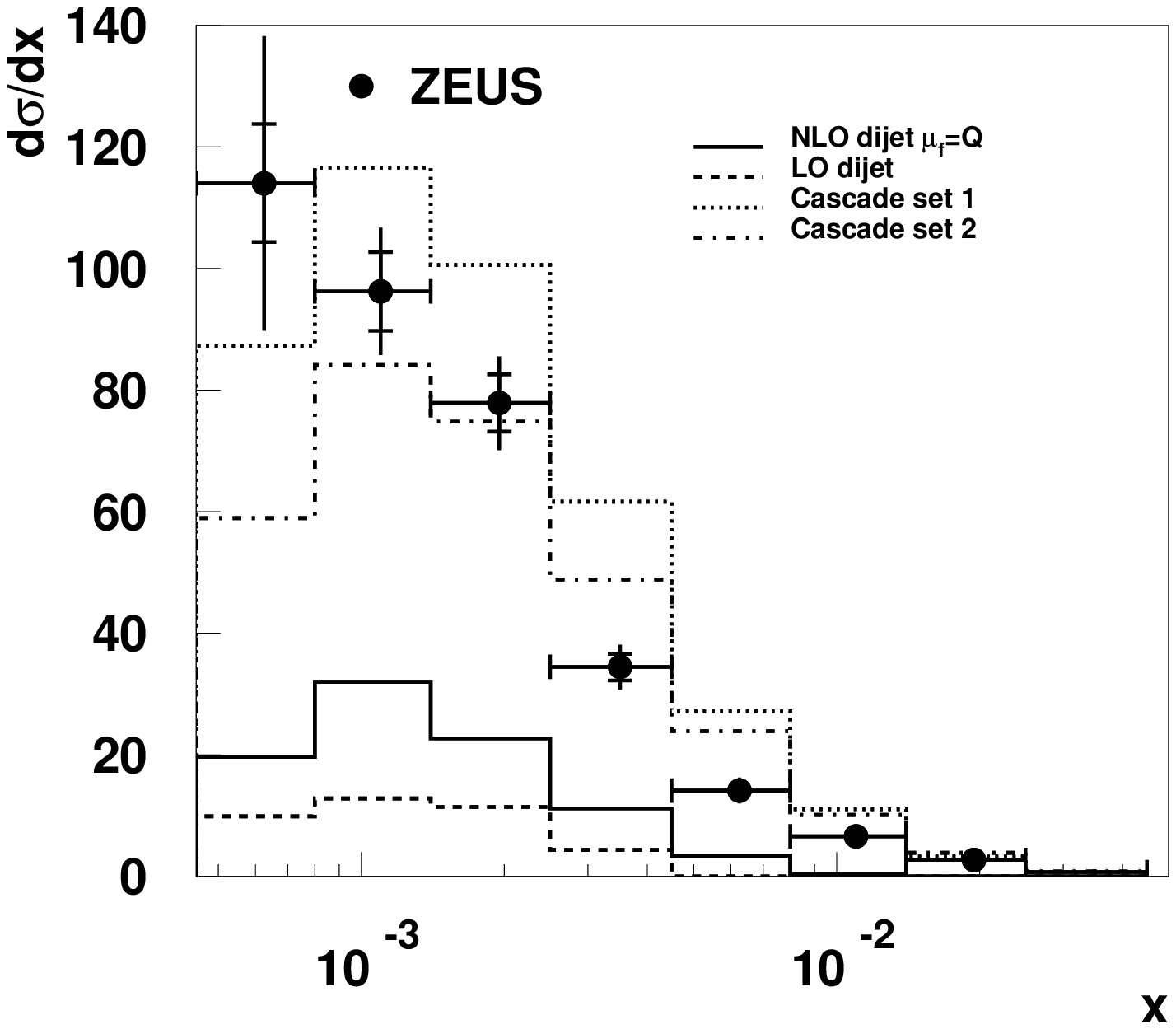,width=6cm}
\end{minipage}
\caption[bla]{Forward jet data from H1 and ZEUS compared to the 
\textsc{Cascade} MC and LO and NLO dijet calculations\cite{forward}.}
\label{forwardjet}
\end{figure}

Results for forward jet production at HERA are shown in 
Fig.~\ref{forwardjet}.
Here we have a comparatively large separation in rapidity, and we see that
LO and NLO dijet calculations are far below the data. 
The CCFM model gives a much better description, but also 
here we see that the best fit is obtained including only the singular 
terms in the splitting function. The reason for this is still not understood;
is there some dynamical mechanism which somehow compensates the effect
of the non-singular terms? 

\subsection{Minimum bias and underlying events in $pp$ collisions}
 
In hadron-hadron collisions collinear factorization works well for 
calculations of high-$p_\perp$ jets. However, in this formalism the minijet 
cross section 
diverges with $\sigma_{jet} \sim 1/p_\perp^4$, which implies that also 
the total $E_\perp$ diverges. This implies the need for a soft cutoff, and 
in \textsc{Pythia} fits to experimental data give a cutoff 
$p_{\perp 0} \sim 2$ GeV. 
This cutoff is also growing with energy, which makes it difficult to 
extrapolate safely to the high energies at LHC.

The symmetry between the two ends of the parton chain implies that the LDC 
formalism also is applicable to hadron-hadron collisions\cite{pp}. 
In the $k_\perp$-\emph{factorization} formalism the off-shell matrix element 
does not blow up when the exchanged transverse 
momentum $k_\perp\!\rightarrow\!0$, and we have therefore a dynamical cutoff 
for soft minijets.
The cross section for a \emph{chain} in $pp$ collisions (which 
possibly may contain more than one hard subcollision) can thus be obtained
from the fit to DIS data. An important point is here that the result is 
insensitive to the soft cutoff, $Q_0$, in the evolution.
DIS data can be fitted with different values for $Q_0$, if the input 
distribution $f_0(x,Q_0^2)$ is adjusted accordingly. If $Q_0$ is increased,
the number of hard chains decreases, but at the same time the number of 
soft chains (for which all emissions have $q_\perp \!< Q_0$) increases, so that
the total number of chains is approximately unchanged.

There are two sources for \emph{multiple interactions}: It is possible to have 
two hard scatterings in the same chain, and there may be more than one chain 
in a single event. The LDC model, when applied to $pp$ collisions, can predict 
the correlations between hard scatterings within one chain, and also the 
average number of chains in a single event. The experimentally observed 
``pedestal effect'' indicates that the hard subcollisions are 
highly correlated, so that central 
collisions have many minijets, while peripheral collisions have fewer 
minijets.
In \textsc{Pythia} comparisons with data favour a distribution in 
the number of subcollisions, which is very close to a geometric 
distribution\cite{pythiasoft}.

\begin{figure}
\begin{minipage}{13cm}
\epsfig{figure=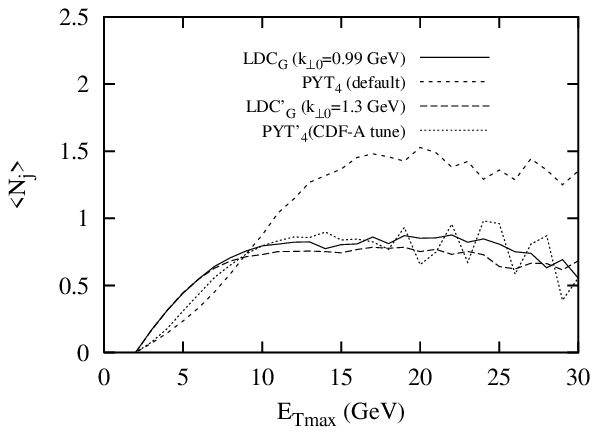,width=6cm}\hspace{-4mm}
\epsfig{figure=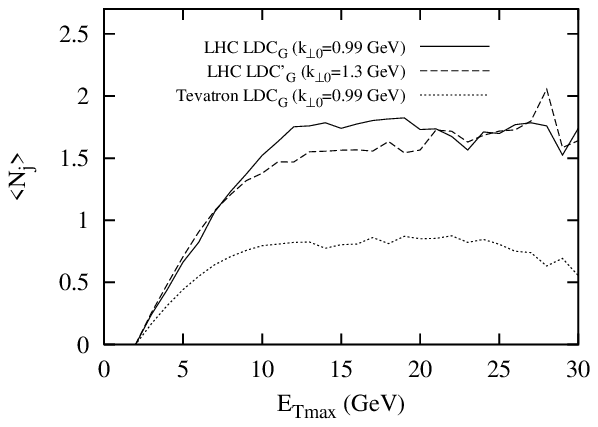,width=6cm}
\end{minipage}
\caption{The average number of minijets in the ``minimum azimuth 
  region'' for $|\eta|<2.5$ \emph{vs}. $E_\perp$ for the 
  hardest jet. \emph{Left}: For $\sqrt s =1.8$ TeV. \emph{Right}: For 14 TeV.}
\label{fig:minijets}

\end{figure}

Some preliminary results from the LDC model are shown in 
fig.~\ref{fig:minijets}. Here a geometric distribution is assumed for 
the number of chains in one event.
 Fig. \ref{fig:minijets}a shows the number of minijets 
in the ``minimum azimuth region'' $60^\circ < \phi < 120^\circ$ 
at $\sqrt s =1.8$ TeV. The two
LDC curves are obtained for soft cut-off values 0.99 and 1.3 GeV, showing the
insensitivity to this cut-off. The two \textsc{Pythia} curves correspond to 
default parameter values, and parameters tuned to CDF data\cite{Field}. We note 
that the LDC result agrees very well with the tuned \textsc{Pythia} result.
Fig. \ref{fig:minijets}b shows corresponding results for LHC. Also
here the two curves correspond to different cut-off values, and for comparison 
the result for 1.8 TeV is also indicated. We see that the activity increases 
by a little more than a factor of 2 between the two energies.

The symmetry of the formalism implies that the chains join at one end at 
the same rate as they multiply at the other. The chain cross section grows 
like $s^\lambda$, and therefore the average chain multiplicity satisfies 
$<\!\!n_{\mathrm{chain}}\!\!>\, \propto s^\lambda/\sigma_{\mathrm{tot}}$. 
Thus the results also may
have implications for unitarization, saturation and diffraction. Work in 
these areas is in progress.

\section{Conclusions}

Our main conclusions can be summarized as follows:

- Unintegrated parton densities are not observables, and their properties 
depend strongly on the definitions.

- Different formalisms give similar results for 
$\mathcal{F}(x,k_\perp^2,\bar{q}^2=4 k_\perp^2)$.
This also implies that the predictions for observable quantities often are
similar.

- The r\^ole of the non-singular terms in the splitting 
functions is still a problem. 

- Observables without a large rapidity separation are often well described 
by higher order matrix elements plus DGLAP evolution.

- There is a close relation between DIS and high energy $pp$ collisions.
The properties of the underlying event and minimum bias events can be predicted
from DIS data.

\section*{Acknowledgements}
I am indebted to Leif L\"onnblad and Hannes Jung for help in the
preparation of this talk.

\end{document}